\documentclass[10pt,conference]{IEEEtran}
\IEEEoverridecommandlockouts

\usepackage{hyperref}

\usepackage[table,xcdraw]{xcolor}
\usepackage{multirow}

\usepackage{soul}
\usepackage{cite}

\usepackage{amsmath,amssymb,amsfonts}
\usepackage{algorithmic}
\usepackage{graphicx}
\usepackage{textcomp}
\usepackage{xcolor}
\def\BibTeX{{\rm B\kern-.05em{\sc i\kern-.025em b}\kern-.08em
    T\kern-.1667em\lower.7ex\hbox{E}\kern-.125emX}}

\newcommand{\newlineauthors}{
  \end{@IEEEauthorhalign}\hfill\mbox{}\par
  \mbox{}\hfill\begin{@IEEEauthorhalign}
}

\newcommand{\myhl}[1]{#1}

\newcolumntype{M}[1]{>{\centering\arraybackslash}m{#1}}

\usepackage{flushend}
\usepackage{comment}
    
\begin{document}

\title{Issue Auto-Assignment in Software Projects with Machine Learning Techniques}

\author{
\IEEEauthorblockN{Pedro Oliveira, Rossana M. C. Andrade, Isaac Barreto}
\IEEEauthorblockA{
  \textit{Group of Computer Networks, Software}         \\ 
  \textit{Engineering and Systems (GREat)}              \\ 
  \textit{Federal University of Cear\'{a} - Brazil}     \\
  \{pedromartins,rossana,isaacbarreto\}@great.ufc.br
}
\and
\IEEEauthorblockN{Tales P. Nogueira}
\IEEEauthorblockA{
  \textit{University of the International}      \\
  \textit{Integration of the Afro-Brazilian}    \\
  \textit{Lusophony (Unilab) - Brazil}          \\
  tales@unilab.edu.br
}
\and
\IEEEauthorblockN{Leandro Morais Bueno}
\IEEEauthorblockA{
\textit{LG Electronics}     \\
\textit{http://lge.com/br}  \\
São Paulo, Brazil           \\
leandro.bueno@lge.com
}
}

\maketitle

\begin{abstract}
Usually, managers or technical leaders in software projects assign issues manually. This task may become more complex as more detailed is the issue description. This complexity can also make the process more prone to errors (misassignments) and time-consuming. In the literature, many studies aim to address this problem by using machine learning strategies. Although there is no specific solution that works for all companies, experience reports are useful to guide the choices in industrial auto-assignment projects. This paper presents an industrial initiative conducted in a global electronics company that aims to minimize the time spent and the errors that can arise in the issue assignment process. As main contributions, we present a literature review, an industrial report comparing different algorithms, and lessons learned during the project.
\end{abstract}


\section{Introduction}
A software development process is a set of related activities that deals with the whole life cycle of software products \cite{sommerville2016software}. During the software development, technical, collaborative, and management activities are performed to specify, design, implement, and test the system. Each of these macro-activities has complex challenges, constituting broad research areas as we can observe in the academic community.

\myhl{One of these challenges is the issue assignment (also known as bug triage). Issues bring a description of tasks or a report of unexpected software behavior, and this challenge is characterized by the need to manage the issues' life cycle from registration until their closing by the development team} \cite{sommerville2016software, aktas2020automated}. 

\myhl{To support this process, it is common to use tracking systems, such as Jira}\footnote{\href{https://www.atlassian.com/software/jira}{Jira website: https://www.atlassian.com/software/jira}}\myhl{, and Redmine}\footnote{\href{https://www.redmine.org}{Redmine website: https://www.redmine.org}}\myhl{. However, for large software projects, the manual assignment of issues is error-prone and time-consuming due to the volume of new records that can be done by clients, stakeholders, technical leaders, internal developers, and testers} \cite{alenezi2013efficient, cavalcanti2014combining}\myhl{. This complexity motivates the development of approaches based on learning algorithms to automate and optimize the issue assignment} \cite{zhang2003machine}.

Since a correct assignment of tasks is crucial for the proper development of a software product, it is necessary to ensure that the most suitable team members will perform these activities. For some activities such as \myhl{solving a very specific bug}, it is essential to have information about the team members' technical knowledge. However, this information is not always available, or there are no significant differences concerning the technical expertise that justifies a member's choice instead of another one. In this scenario, historical data about previously assigned issues can be used by learning systems to create decision support systems.

In this context, the issue assignment problem --- which is a subtype of a combinatorial problem --- can be formally defined as \cite{SALMAN2002363}: given two sets $A = \{a_{1},..., a_{|A|}\}$ for assignees and $I = \{i_{1},..., i_{|I|}\}$ for issues, find a function $f: A \rightarrow I$ or a machine learning model that minimizes the cost function ($C: A \times I \rightarrow \mathbb{R}$) \myhl{generally} described by Equation \ref{eq_assign}.

\begin{equation}
    \label{eq_assign}
    \sum_{a \epsilon A}^{} C (a, f(a))
\end{equation}

\begin{table*}[!t]
\centering
\caption{Papers selected in the literature review compared with our proposal.}
\label{tab:rworks}
\scalebox{0.78}{
{\def\arraystretch{1.3}
\begin{tabular}{|c|M{1.2cm}|m{2.5cm}|m{3.6cm}|m{4.5cm}|m{4cm}|}
\hline
\rowcolor[HTML]{EFEFEF}
\textbf{Work} & \textbf{Issue type} & \textbf{Techniques} & \textbf{Features} & \textbf{Machine learning support tools} & \textbf{Datasets} \\ \hline
\myhl{Helming et al.} \cite{helming2010automatic} & Bug and general tasks & kNN, Decison trees, SVM, and Naive Bayes & Work item description & Java Data Mining Package, WEKA, LIBLINEAR, and MALLET & Three projects located in UNICASE tool \\ \hline
\myhl{Aljarah et al.} \cite{aljarah2011selecting} & Bug & Bayesian Network Classifier & Bug report & WEKA & Eclipse projects: Core Component, UI Component, SWT Component \\ \hline
\myhl{Sureka} \cite{sureka2012learning} & Bug & TF-IDF and DLM models & Bug report (title and description) & LingPipe & Eclipse and Mozilla projects \\ \hline
\myhl{Alenezi et al.} \cite{alenezi2013efficient} & Bug & Naive Bayes & Bug report & WEKA & Eclipse-SWT, Eclipse-UI, NetBeans, and Maemo projects \\ \hline
\myhl{Cavalcanti et al.} \cite{cavalcanti2014combining} & Change Request & Rule-based Expert System, Information Retrieval, and SVM & Change request, severity, component & Drools and WEKA & One project at Brazilian Federal Data Processing Service (SERPRO) \\ \hline
\myhl{Jonsson et al.} \cite{jonsson2016automated}  & Bug & Stacked Generalization ensemble & Title and description, Submitter type, Submitter Site, priority & WEKA & One project at an Automation company and four projects at a Telecom company \\ \hline
\myhl{Ded{\'\i}k and Rossi} \cite{dedik2016automated}  & Bug & SVM + TF-IDF & Bug report & Not identified & Proprietary dataset and Firefox project \\ \hline
\myhl{Sharma et al.} \cite{sharma2017reduction} & Bug & Apriori + Kmeans & Severity, priority, component, and OS & MATLAB and Rapid Miner & Seamonkey, Firefox, and Bugzilla projects \\ \hline
\myhl{Peng et al.} \cite{peng2017improving} & Bug & Relevant search techniques & Bug report & Not identified & Mozilla and Eclipse projects \\ \hline
\myhl{Hern{\'a}ndez-Gonz{\'a}lez et al.} \cite{hernandez2018learning} & Defect Report & Bayesian Network Classifiers & Summary, description, severity & Own implementation & Compendium and Mozilla projects \\ \hline
\myhl{Choquette-Choo et al.} \cite{choquette2019multi} & Bug & Deep Neural Network & Bug report & Not identified & Google Chromium project \\ \hline
\myhl{Lee and Seo} \cite{lee2020improving} & Bug & LDA & Bug report & NLP Python libraries & Bugzilla and Android bug reports \\ \hline
\rowcolor[HTML]{DAE8FC}
Our work  & Bug & SGDText, Logistic Regression, and Random Forest & Summary and description & WEKA & Proprietary (LGE Brazil) dataset \\ \hline
\end{tabular}}}
\end{table*}

The cost function can describe different restrictions based on the company context. For example, it is possible to adapt this function considering each assignee's relevant experience to solve the new issue \cite{alenezi2013efficient}, the operation cost to assign an issue for a specific developer \cite{kashiwa2020release}, or even the average time required to solve similar issues \cite{jeong2009improving}. \myhl{In our case, the cost function was modeled to reduce misassignments as described following.}

Thus, this paper presents a report based on our experience in a project whose main goal was to minimize the time spent during the issue assignment process as well as the number of errors (misassignments). In this case, a new issue should be initially assigned to one of the technical leaders taking into account their responsibilities. Thus, we consider a misassignment if this first assignment is wrong. Finally, the experiments took place in a setting of real projects of the LG Electronics mobile division in São Paulo, Brazil (LGESP).

LGE Brazil\footnote{\href{https://www.lg.com/br}{LGE Brazil website: https://www.lg.com/br}}, which is part of the LG group, is one of Brazil's largest electronics companies, operating in this country for over 25 years. It has an office in São Paulo and two production units in Manaus and Taubaté cities. For this project, we established contact with the mobile division located in São Paulo. This division is responsible, among other activities, for solving issues related to the company's products. Due to non-disclosure constraints, some confidential information was omitted or anonymized without prejudice to our process and results' overall understanding.

Our main contributions in a nutshell are:

\begin{itemize}
    \item a literature review presenting the most recent studies addressing the issue assignment problem in academic and industrial environments;
    \item a comparison among different algorithms and strategies to tackle the manual issue assignment problem in a large company real setting; and 
    \item lessons learned in partnership with LGESP, which can help other researchers and practitioners.
\end{itemize}

\section{Related Work}
\label{sec:rWorks}
We performed a literature review on papers published in the last ten years (2010 - 2020). We used as search string the following terms: ``\textit{{(software engineering}) AND ({data mining} OR {machine learning}) AND (issue OR task OR bug OR \{defect report\} OR \{trouble report\}) AND (assignment OR attribution OR allocation)}''. For the search, Elsevier's Scopus was used as the data source.

Initially, 65 papers were recovered, but only twelve (12) were chosen after reading the title and abstract. The eligibility criteria were: be a primary study, written in English, fully available on the Internet, and with more than 4 pages; be published in conferences or journals; and discuss processes, methods, or tools to tackle the issue assignment problem.

The date range (from 2010 to 2020) was defined to cover recent initiatives regarding this problem, and the Scopus database was selected based on its coverage\footnote{Scopus Content Coverage Guide: \href{https://www.elsevier.com/?a=69451}{https://www.elsevier.com/?a=69451}.} of software engineering venues and relevant digital libraries such as ACM, IEEExplorer, Science Direct, and Springer. Thus, the selected papers represent a suitable sample to describe this study area.

Table \ref{tab:rworks} summarizes the twelve selected papers and our work considering \myhl{five aspects: the issue type, techniques, features, machine learning supporting tools, and datasets used in each of the works}. With this summary, it is possible to observe that most studies used open-source projects to validate their proposals. Also, the issue assignment has been modeled as a textual classification focused on bug triage.

The analysis of these papers highlights that the issue auto-assignment problem was not entirely overcome, and there is no ``silver bullet''. The context of each project has a significant influence on the results obtained by machine learning techniques. Hence, it is essential to conduct applied studies that result in experience reports with lessons learned. These lessons can guide researchers and practitioners in applying the most appropriate strategies for their context. In this work, we present our experience report of the research conducted in a large electronics company.

\section{Our approach}
\label{sec:method}
As pointed out in Section \ref{sec:rWorks}, although the issue assignment problem has been studied and reported in the literature, there is no silver bullet solution once the approaches used to tackle this problem depend on the context. Hence, experience reports are useful to guide industrial projects' choices, and this section presents the approach adopted to study the issue assignment problem in LGESP.

As a starting point, we decided to use the CRISP-DM (CRoss Industry Standard Process for Data Mining) process \cite{wirth2000crisp}, which was created for supporting researchers and practitioners in the execution of data mining industry projects. This decision was based on its characteristics that allow iterative conduction resulting in artifacts to be deployed and used by the company. Next, we describe the six phases of CRISP-DM and how we had conducted each of them.

\subsection{Business Understanding}

The first phase focuses on business understanding to clarify the fundamental requirements and objectives. This phase is quite essential to create a suitable plan for the project. In our case, we have done meetings with managers and team leaders to understand the problem's details and how it impacts the company's work. Also, it was possible to establish a quick communication channel between the managers and the research team.

As a result of this phase, we have built a project plan describing the problem context, the goals, and the activity schedule. Regarding the problem context, we have identified that issues should initially be assigned to the leader of six different sub-teams ($ST_{1}, ST_{2}, ST_{3}, ST_{4}, ST_{5},S T_{6}$) and these sub-teams were organized into two teams $T_{A} = \{ST_{1}, ST_{2}, ST_{3}\}$, and $T_{B} = \{ST_{4}, ST_{5}, ST_{6}\}$. Sub-teams have different responsibilities regarding bugs, \textit{e.g.}, a sub-team solves troubleshooting related to networks and protocols. In this way, leaders are responsible for distributing issues among specialists and engineers. This information is essential to design the auto-assignment models\footnote{From this point, whenever a sub-team assignment is mentioned, we refer to the assignment to the leader of this sub-team.}.

\subsection{Data Understanding}

After business understanding, we analyzed the historical data -- acquired from the company's issue tracking system -- aiming for initial insights and identifying the relationship among data attributes and issues' assignments.

The raw data included the following attributes: \textit{key} (an unique identifier), \textit{summary} (the issue's subject), \textit{assignee} (the user who was manually assigned to fix the issue), \textit{reporter} (the user that reported the issue), \textit{components} (which software components were affected), \textit{priority} (values ranged from P0 to P3), \textit{attach \#} (number of files attached to the issue), \textit{created} (date of creation), \textit{updated}  (date of last update), \textit{due date}, \textit{labels} (some user defined labels, similar to \textit{components}), and \textit{description} (the issue's body text).

We have created a set of data visualizations using Tableau\footnote{Tableau website: \href{https://www.tableau.com}{https://www.tableau.com}.}. This exploratory data analysis helped us to understand the type and priority of issues, the evolution of the status, and the issue creation frequency. Then, we conducted meetings to discuss our observations, such as the correlation - \myhl{visually observed from the data distribution} - between some attributes and the final assignment (\textit{e.g.}, component and priority). According to the experts, this correlation occurs because some attributes are defined after the assignment and during the issue investigation. In this phase, it was possible to improve the business knowledge, and it became clear that the machine learning models in our approach could use only textual attributes (summary and description), which are present since the issue is first reported.

\subsection{Data Preparation}

Regarding data preparation, CRISP-DM recommends the execution of many activities to build the final dataset. These activities include recovering raw data, performing attribute selection, data cleaning, creating new attributes, and transforming the dataset to be used in modeling tools.

\begin{table}[h]
\centering
\caption{Data distribution considering teams and sub-teams.}
\label{tab:datadist}
\scalebox{1}{
{\def\arraystretch{1.3}
\begin{tabular}{|c|c|r|}
\hline
\rowcolor[HTML]{EFEFEF} 
\textbf{Teams} & \textbf{Sub-team} & \textbf{Instances} \\ \hline
 & $ST_{1}$ & 1,160 \\ \cline{2-3} 
 & $ST_{2}$ & 752 \\ \cline{2-3} 
\multirow{-3}{*}{$T_{A}$} & $ST_{3}$ & 310 \\ \hline
 & $ST_{4}$ & 1,691 \\ \cline{2-3} 
 & $ST_{5}$ & 1,363 \\ \cline{2-3} 
\multirow{-3}{*}{$T_{B}$} & $ST_{6}$ & 408 \\ \hline
\end{tabular}}}
\end{table}

In this work, we have initially retrieved 8,344 issues from January 2018 to August 2020. This period was chosen because several internal changes happened before 2018 that could affect the results' quality. Then, we removed not closed, duplicated, and unassigned issues. After this preprocessing, 5,684 useful issues remained in the dataset. Table \ref{tab:datadist} presents the distribution by teams and sub-teams.

As discussed in the \textit{Business Understanding} and in the \textit{Data Understanding} subsections, we have faced the challenge of performing an automatic issue assignment for six sub-team leaders using only textual data to build the machine learning models (\myhl{\textit{i.e.}, we need to predict the correct sub-team for the new issues from their report}). Thus, it was necessary to apply natural language processing (NLP) techniques to the unstructured dataset (subject and description). First, we removed special characters and unwanted words (\textit{e.g.}, HTML tags, accentuation, decimal, and hexadecimal numbers, punctuation, and words with less than three characters). Then, the subject and description were joined. After, we turned the resulting text into tokens, removed stopwords, and applied the stemming algorithm \cite{thomas2014mining}. 

In this work, we used a set of stopwords provided by the Rainbow project \cite{McCallumLibbow}, and the stemming was based on the Lovins stemmer \cite{lovins1968development}. They were chosen due to their availability on WEKA and the ability to remove noise, improving the classifiers' performance.

To conclude the data preparation, we used the StringToWordVector algorithm to convert the final text attribute into a set of features representing word frequency considering the TF-IDF algorithm \cite{robertson2004understanding} and the InfoGain algorithm to remove features that do not add information about the assignment.

\subsection{Modeling}

In the modeling phase, nine different classifiers were selected and applied to find the most suitable one to tackle the problem. This selection was empirical and exploratory, aiming to cover classifiers with varying characteristics and considering the ones most used in the literature. The selected classifiers were Naïve Bayes, SVM (Support Vector Machines) provided by LibSVM library, J48, Random Forest, kNN (k-Nearest Neighbours), Logistic Regression, Naïve Bayes Multinomial Text, SGDText (Stochastic Gradient Descent adapted to deal with textual data), and Zero Rule (ZeroR), which is a simple classifier that chooses the majority class as the classification for all instances (used as a baseline) \cite{John1995,Chang2001,Quinlan1993,Breiman2001,Aha1991,Landwehr2005,su2008discriminative,zhang2004solving,de2006mining}.

The modeling of these classifiers was performed using WEKA - a machine learning workbench\footnote{\href{https://www.cs.waikato.ac.nz/ml/weka/index.html}{WEKA website: https://www.cs.waikato.ac.nz/ml/weka/index.html}} \cite{hall2009weka}. In addition to being quite used in other works (see Table \ref{tab:rworks}), this tool enabled us to quickly build and evaluate several models.

\subsection{Evaluation}

The following questions guided the evaluation:

\begin{itemize}
    \item \textbf{RQ1}: considering the company context, what is the suitable strategy (S1: classify the issues in just one step or S2: classify the issues in two steps) to perform the issue assignment?
    
    \item \textbf{RQ2}: which are the best classifiers to do this assignment?
    
    \item \textbf{RQ3}: how to deal with the imbalanced classes?
    
    \item \textbf{RQ4}: considering that new mobile technologies arise and others are no longer used; or even that the issues' writing patterns changes over time according to new internal guidelines or process modifications, how often should models be retrained to maintain a high accuracy?
    
    \item \textbf{RQ5}: how much effort is saved by the auto-assignment when compared to the manual issue assignment?
\end{itemize}

Regarding the metrics used to analyze these RQs, we adopted accuracy, the harmonic mean of precision and recall, called F-Measure \cite{ian2005data}, and the average time spent in the assignment of one issue (measured in seconds). At this point, it is important to highlight that we did not use other common metrics such as MRR (Mean Reciprocal Rank) or top-k accuracy \cite{sajedi2020guidelines} because, in our case, a top-k recommendation does not bring value for the team leaders. 

\begin{table}[h]
\centering
\caption{Strategies analyzed in our evaluation.}
\label{tab:strategies}
\scalebox{0.82}{
{\def\arraystretch{1.3}
\begin{tabular}{|c|l|c|}
\hline
\rowcolor[HTML]{EFEFEF} 
\textbf{RQ} & \textbf{Strategy} & \multicolumn{1}{l|}{\cellcolor[HTML]{EFEFEF}\textbf{Experiment detail}} \\ \hline
 & S1: considering the six sub-team leaders & \multicolumn{1}{l|}{E1: all classifiers} \\ \cline{2-3} 
 &  & \multicolumn{1}{l|}{E2: all classifiers (teams)} \\ \cline{3-3} 
 &  & \multicolumn{1}{l|}{E3: all classifiers ($T_{A}$ sub-teams)} \\ \cline{3-3} 
\multirow{-4}{*}{\begin{tabular}[c]{@{}c@{}}RQ1\\ RQ2\end{tabular}} & \multirow{-3}{*}{S2: classify the issues in two steps} & \multicolumn{1}{l|}{E4: all classifiers ($T_{B}$ sub-teams)} \\ \hline
 & S3: undersampling &  \\ \cline{2-2}
 & S4: oversampling &  \\ \cline{2-2}
\multirow{-3}{*}{RQ3} & S5: SMOTE &  \\ \cline{1-2}
RQ4 & S6: sliding windows &  \\\cline{1-2}
RQ5 & S7: interview with the managers & \multirow{-5}{*}{\begin{tabular}[c]{@{}c@{}}SGDText, \\ Logistic Regression,\\ and Random Forest\end{tabular}} \\ \hline
\end{tabular}}}
\end{table}

Table \ref{tab:strategies} shows more details about the strategies used for answering each research question. To answer RQ1 and RQ2, all classifiers were evaluated, while for RQ3 and RQ4, only the best classifiers were used. In the case of RQ5, due to some restrictions, it was not possible to perform a more formal empirical evaluation (\textit{e.g.}, controlled experiment, or case study). Thus, we collected from domain experts the estimated effort spent on this activity to compare with the results achieved with our proposal. Section \ref{sec:experiments} presents more details about the evaluation.

\subsection{Deployment}

The last phase is deployment. At this point, it is essential to take into account how the created models will be used.

In this project, it was decided that models should be made available as a Web service to facilitate the integration with the company's issue tracking system. It was developed over the Docker container platform \cite{merkel2014docker} with three main modules: an API following the REST architecture style \cite{wilde2011rest},  a Classification Service, and a Training Service. By default, the results generated by the Classification Service follow the JSON format. However, it is possible to export the results to CSV (comma-separated values) format, which is especially useful for batch classification.

Regarding Training Service, after a training round, the serialized model is stored in a Cloud bucket, and the performance metrics are kept in a NoSQL database. In this way, the user can choose the models to be used in the Classification Service. In addition to the core services and API, the following features were developed to support the evolution of the whole process:

\begin{itemize}
    \item \textbf{Training center and report}: with this feature, it is possible to start a new training of the best models, analyze reports with training performance metrics, and decide the suitable models to use in the Classification Service.
    
    \item \textbf{Batch classification}: with this feature, the manager can inspect the assignment service submitting a batch of issues, and getting a CSV file as a result. This file brings the issues data (key, summary, and description) and the team and sub-team classification result.
\end{itemize}

\section{Experiments}
\label{sec:experiments}
\begin{figure}[b]
    \centering
    \includegraphics[width=0.6\columnwidth]{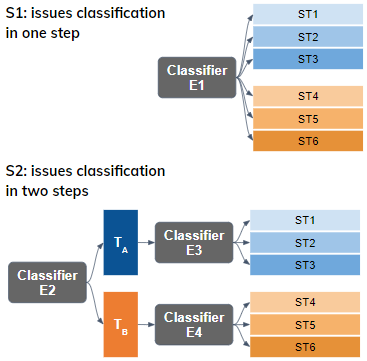}
    \caption{A graphical representation of S1 and S2 strategies.}
    \label{fig:s1_s2}
\end{figure}

RQ1 and RQ2 guided the investigation to find a suitable strategy and the best classifier to perform the issue assignment. Thus, four experiments were carried out (E1, E2, E3, and E4) in WEKA, with ten executions for each classifier and 10-fold cross-validation. Figure \ref{fig:s1_s2} shows a graphical representation of these strategies. The first one (S1) defines the classification in just one step to correctly define the issue sub-team. In contrast, the second strategy (S2) performs a classification in two steps, seeking, first to define which team each issue should be forwarded to and then to which sub-team. As S2 uses a classifier chain, we used Equation (\ref{eq:acc_s2}) to calculate its final accuracy, which gives the probability of correct predictions using the best classifiers in experiments E2, E3, and E4. \myhl{This equation was defined based on the conditional probability that an issue belongs to a specific team/sub-team, and the classifier returns the correct prediction} \cite{jain2008art}.

\begin{equation}
    \centering
    Acc(S2) = [Acc(E2) \times 0.5] \times [Acc(E3) + Acc(E4)]
    \label{eq:acc_s2}
\end{equation}

After these four experiments, three resample strategies were evaluated: undersampling (S3), oversampling (S4), and SMOTE (S5) \cite{al2002, chawla2002smote, ian2005data}. From this stage, all experiments were carried out with the best classifiers of the previous experiments. For S3 and S4, \textit{SpreadSubsample} and \textit{Supervised Resample} filters were used, respectively. In S4, the sample size was two times the percentage of data that belonged to the majority class. With SMOTE (\textit{Synthetic Minority Oversampling TEchnique}), synthetic instances were created from the relationship between the number of instances in the majority class and the number of instances in the minority classes.

For answering RQ4, we sought to understand the practical challenges of using our approach to support the issue assignment. To do this, we analyzed how models' accuracy decreased over time and used a sliding window to identify the most suitable frequency to repeat the training task. For this experiment, we used six months, one year, and two years as the training windows, and one week, one month, six months, and one year as the testing windows.

As previously mentioned, all experiments were supported by WEKA. The statistical analyses were performed in this tool's experimentation environment, using the corrected paired t-test \cite{nadeau2000inference} with a significance level of 0.05.

Finally, for RQ5, we interviewed domain experts to collect data concerning the time spent with the manual issue assignment. This interview occurred in one of the project meetings, and it was focused on two questions: i) what is the frequency of this activity? ii) how much time is spent on it? Then, the time was converted to seconds to compare with the time spent using the automated assignment.

\section{Results and Discussion}
\label{sec:results}
This section brings the results obtained in our experiments and provides a discussion about them.

\begin{table*}[]
\centering
\caption{Results of the experiments conducted for RQ1 and RQ2.}
\scalebox{0.75}{
{\def\arraystretch{1.3}
\begin{tabular}{|l|c|c|c|c|c|c|c|c|}
\hline

\rowcolor[HTML]{EFEFEF} 
\cellcolor[HTML]{EFEFEF} & \multicolumn{2}{c|}{\textbf{S1}} & \multicolumn{6}{c|}{\textbf{S2}} \\ \cline{2-9} 
\rowcolor[HTML]{EFEFEF}
\cellcolor[HTML]{EFEFEF} & \multicolumn{2}{c|}{\textbf{E1}} & \multicolumn{2}{c|}{\textbf{E2}} & \multicolumn{2}{c|}{\textbf{E3}} & \multicolumn{2}{c|}{\textbf{E4}} \\ \cline{2-9} 

\multirow{-3}{*}{\cellcolor[HTML]{EFEFEF}\textbf{Classifier}} & Accuracy & F-Measure & Accuracy & F-Measure & Accuracy & F-Measure & Accuracy & F-Measure \\ \hline
ZeroR               & 29.75(0.05)   & - & 60.91(0.06) & - & 52.21(0.09) & - & 48.84(0.09) & - \\ \hline
Naive Bayes         & 56.68(1.71)   & 0.55(0.02) & 77.23(1.76) & 0.77(0.02) & 83.85(2.45) & 0.84(0.02) & 56.66(2.75) & 0.57(0.03)\\ \hline
SVM                 & 75.55(1.60)   & 0.76(0.02) & 94.17(0.91) & 0.94(0.01) & 92.76(1.47) & 0.93(0.01) & 67.98(2.65) & 0.68(0.03)\\ \hline
J48                 & 70.24(1.85)   & 0.70(0.02) & 92.06(1.04) & 0.92(0.01) & 89.35(1.93) & 0.89(0.02) & 66.99(2.47) & 0.67(0.02)\\ \hline
Random Forest       & 76.03(1.60) * & 0.76(0.02) * & 94.35(0.85) & 0.94(0.01) & 92.07(1.85) & 0.92(0.02) & \cellcolor[HTML]{DAE8FC}74.13(2.38) & \cellcolor[HTML]{DAE8FC}0.74(0.02)\\ \hline
kNN                 & 73.34(1.92)   & 0.73(0.02) & 92.37(1.09) & 0.92(0.01) & 88.41(1.85) & 0.88(0.02) & 71.73(2.34) & 0.72(0.02)\\ \hline
Logistic Regression & \cellcolor[HTML]{DAE8FC}76.83(1.46) & \cellcolor[HTML]{DAE8FC}0.77(0.01) & 95.20(0.80) & 0.95(0.01) & \cellcolor[HTML]{DAE8FC}93.95(1.47) & \cellcolor[HTML]{DAE8FC}0.94(0.01) & 70.17(2.34) & 0.70(0.02)\\ \hline
Naive Bayes MT      & 29.75(0.05) & - & 60.91(0.06) & - & 52.21(0.09) & - & 48.84(0.09) & - \\ \hline
SGD Text            & - & - & \cellcolor[HTML]{DAE8FC}97.13(0.63) & \cellcolor[HTML]{DAE8FC}0.97(0.01) & - & - & - & - \\ \hline
\end{tabular}}}
\label{tab:result_part_one}
\end{table*}

\begin{table*}[h]
\centering
\caption{Results of the experiments concerning the resample strategies.}
\scalebox{0.75}{
{\def\arraystretch{1.3}
\begin{tabular}{|p{4.2cm}|c|c|c|c|c|c|}
\hline
\rowcolor[HTML]{EFEFEF} 
\cellcolor[HTML]{EFEFEF} & \multicolumn{2}{c|}{\textbf{SGDText (E2)}} & \multicolumn{2}{c|}{\textbf{Logistic Regression (E3)}} & \multicolumn{2}{c|}{\textbf{Random Forest (E4)}} \\ \cline{2-7}
\multirow{-2}{*}{\cellcolor[HTML]{EFEFEF}\textbf{Resample strategies}} & Accuracy & F-Measure & Accuracy & F-Measure & Accuracy & F-Measure \\ \hline
Without resampling & 97.13(0.63) * & 0.97(0.01) * & \cellcolor[HTML]{DAE8FC}93.95(1.47) & \cellcolor[HTML]{DAE8FC}0.94(0.01)  & \cellcolor[HTML]{DAE8FC}74.13(2.38) & \cellcolor[HTML]{DAE8FC}0.74(0.02)\\ \hline
Undersampling & 96.56(0.70) & 0.97(0.01) & 89.61(2.20) & 0.90(0.02)  & 63.81(2.49) & 0.64(0.02)\\ \hline
Oversampling & 96.45(0.69) & 0.96(0.01) & 92.32(1.46) & 0.92(0.01) & 70.96(2.74) & 0.71(0.03)\\ \hline
SMOTE & \cellcolor[HTML]{DAE8FC}97.16(0.65) & \cellcolor[HTML]{DAE8FC} 0.97(0.01) & 93.75(1.58) * & 0.94(0.02) * & 73.84(2.40) * & 0.73(0.02) * \\ \hline
\end{tabular}}}
\label{tab:result_resampling}
\end{table*}

\begin{table*}[]
\centering
\caption{Results of the experiments using sliding windows.}
\scalebox{0.75}{
{\def\arraystretch{1.3}
\begin{tabular}{|c|l|c|c|c|c|c|c|}
\hline
\rowcolor[HTML]{EFEFEF} 
\cellcolor[HTML]{EFEFEF} & \cellcolor[HTML]{EFEFEF} & \multicolumn{2}{c|}{\cellcolor[HTML]{EFEFEF}\textbf{SGDText (E2)}} & \multicolumn{2}{c|}{\cellcolor[HTML]{EFEFEF}\textbf{Logistic Regression (E3)}} & \multicolumn{2}{c|}{\cellcolor[HTML]{EFEFEF}\textbf{Random Forest (E4)}} \\ \cline{3-8} 

\multirow{-2}{*}{\cellcolor[HTML]{EFEFEF}\textbf{Training Window}} & \multirow{-2}{*}{\cellcolor[HTML]{EFEFEF}\textbf{Testing Window}} & Accuracy & F-Measure & Accuracy & F-Measure & Accuracy & F-Measure \\ \hline
 & One Week (51) & \cellcolor[HTML]{DAE8FC}94.51(5.69) & \cellcolor[HTML]{DAE8FC}0.9456(0.06) & 85.98(14.01) & 0.8602(0.14) & 65.17(17.52) & 0.6478(0.19)\\ \cline{2-8} 
 & One Month (16) & 90.65(5.90) & 0.9072(0.06) & 83.47(8.17) & 0.8324(0.09) & 59.58(9.08) & 0.5792(0.09)\\ \cline{2-8} 
\multirow{-3}{*}{One Semester (70)} & One Semester (3) & 82.05(5.88) & 0.8291(0.05) & 84.43(3.06) & 0.8445(0.03) & 61.99(13.04) & 0.5937(0.15)\\ \hline

 & One Week (9) & 92.81(8.10) & 0.9229(0.09) & 87.53(11.29) & 0.8678(0.12) & 52.33(14.77) & 0.5296(0.14)\\ \cline{2-8} 
 & One Month (10) & 87.52(8.58) & 0.8830(0.07) & 85.81(3.69) & 0.8527(0.05) & 57.00(7.82) & 0.5663(0.12)\\ \cline{2-8} 
 & One Semester (4) & 83.46(4.97) & 0.8516(0.03) & 86.98(3.02) & 0.8676(0.03) & 56.39(10.03) & 0.5474(0.17)\\ \cline{2-8} 
\multirow{-4}{*}{One Year (25)} & One Year (2) & 84.46(2.33) & 0.8454(0.02) & 82.08(5.88) & 0.8157(0.06) & 45.97(3.60) & 0.3926(0.05)\\ \hline

 & One Week (15) & 93.54(7.22) & 0.9357(0.07) & \cellcolor[HTML]{DAE8FC}93.17(12.89) & \cellcolor[HTML]{DAE8FC}0.9493(0.09) & \cellcolor[HTML]{DAE8FC}70.12(24.78) & \cellcolor[HTML]{DAE8FC}0.7261(0.23)\\ \cline{2-8} 
 & One Month (5) & 89.27(6.11) & 0.8939(0.06) & 89.87(2.31) & 0.8979(0.02) & 51.57(7.49) & 0.4853(0.11)\\ \cline{2-8} 
 & One Semester (1) & 85.59(0.00) & 0.8599(0.00) & 89.44(0.00) & 0.8947(0.00) & 39.46(0.00) & 0.3130(0.00)\\ \cline{2-8} 
\multirow{-4}{*}{Two Years (22)} & One Year (1) & 86.64(0.00) & 0.8678(0.00) & 88.98(0.00) & 0.8908(0.00) & 39.46(0.00) & 0.3130(0.00)\\ \hline
\end{tabular}}}
\label{tab:result_sliding}
\end{table*}

First, it was necessary to identify a suitable strategy to perform the issue assignment. This point is relevant because it was identified in the company's process that it is crucial to have an assignment service with high accuracy in choosing the team ($T_{A}$ or $T_{B}$). Thus, two strategies were analyzed: S1 and S2. Table \ref{tab:result_part_one} shows the average accuracy and the F-Measure for each experiment. The values between parentheses represent the standard deviation. The asterisk ($*$) identifies values with no statistical difference compared to the best results highlighted in blue. Furthermore, the dash (-) identifies measures that could not be calculated, \textit{e.g.}, in WEKA, \myhl{the SGDText classifier only deals with binary problems. Hence, it was used exclusively in the E2 experiment}. All other tables in this work follow this convention.

For S1, the best result was obtained by Logistic Regression with 76.83\% accuracy, and F-Measure equals 0.77. This was the only experiment in which another classifier with a statistically similar result was identified, namely the Random Forest with 76.03\% accuracy and F-Measure equals 0.76.

To calculate the accuracy for S2, we used Equation (\ref{eq:acc_s2}) using the best results obtained in E2, E3, and E4. In E2, the models were trained to classify the correct team for each issue and the best result was reached by SGDText with 97,13\% of accuracy and F-Measure 0.97. In E3 and E4, the objective was to classify according to sub-teams $ST_{1}, ST_{2}, ST_{3}$, and $ST_{4}, ST_{5},S T_{6}$, respectively. The best results were achieved by Logistic Regression and Random Forest with 93.95\% and 74.13\% of accuracy, respectively. Applying (\ref{eq:acc_s2}), we have: 

\begin{equation}
    \centering
    Acc(S2) = [0.9713 \times 0.5] \times [0.9395 + 0.7413] \rightarrow 0.8163
    \label{eq:finalAcc}
\end{equation}

Thus, it is possible to answer the RQ1 stating that, in our case, it is more accurate (81.63\%) to conduct the classification in two steps (S2), considering the teams' internal organization. This strategy also brings more value to the company due to the high success rate in team classification (97.13\% with SGDText). For RQ2, the classifiers were the SGDText, Logistic Regression, and Random Forest. It is possible to find some works in the literature that also obtained good results with Logistic Regression, Random Forest, and algorithms based on Support Vector Machines \cite{helming2010automatic, cavalcanti2014combining, jonsson2016automated, dedik2016automated}. However, we did not find any work that used SGDText. This classifier implements a Stochastic Gradient Descent and operates directly on textual attributes. This adaptation to work specifically with textual data can explain the good achieved results.

For the SGDText classifier, the best results were obtained by training with a window of one semester and executing a new training every week. In contrast, for the Logistic Regression and Random Forest classifiers, the best configuration consisted of training with data from the last two years and retraining weekly. Here, it is important to highlight three points:
\begin{itemize}
    \item the classifiers' performance is negatively influenced by the data seasonality, \textit{i.e.}, for some periods, there is a large number of issues for team $T_{A}$ compared to $T_{B}$;
    \item the models achieved better results with weekly retraining, probably due to the low number of new issues per week. On average, there are 40 new issues per week; and
    \item the classifications regarding teams (made by SGDText) and sub-teams $ST_{1}, ST_{2}, ST_{3}$ (Logistic Regression) got accuracy greater than 82\% for all window combinations.
\end{itemize}

After choosing the best strategy and the best classifiers, we conducted experiments to evaluate methods that dealt with class imbalance. At this point, three well-known strategies were analyzed: undersampling, oversampling, and SMOTE. The results, presented in Table \ref{tab:result_resampling}, show that none of these strategies improved the accuracy and F-Measure. The results without resampling and with SMOTE were statistically similar, and the other methods showed statistically inferior results. One reason for this is that these methods were not developed to handle textual data even when transformed into word frequency features. Thus, the imbalance problem needs further investigation, and the answer to RQ3 is that the most common resampling methods are not suitable for text classification.

Regarding RQ4, the models' deterioration was analyzed in order to understand the impacts of the changes in the writing of the issues. Experiments were performed considering sliding windows to find a suitable size for collecting training data (training window) and the retraining frequency (test window). The results of these experiments are presented in Table \ref{tab:result_sliding}. The numbers in parentheses in the columns ``Training Window'' and ``Testing Window'' represent the data points used to calculate the metrics. The division does not always strictly follow the temporal pattern because, in some cuttings, there was not enough data for training (due to seasonality in the record of issues), and the learning process failed.

So, the most significant difficulty relies on the $ST_ {4}, ST_ {5}, ST_ {6}$ classification (made by Random Forest), probably due to the similarity of the issues attributed to these sub-teams. Thus, to keep hit rates at the same level found in the cross-validation experiment (Table \ref{tab:result_part_one}), the answer to RQ4 is that the models should be trained using data of the last two years and with weekly retraining.

Finally, for RQ5, we asked domain experts to estimate the time spent on assignments and how often this activity is performed. They replied that this activity is performed daily with an average time of 2 hours and 40 minutes (160 minutes). Considering 8,344 issues registered between January 2018 and August 2020 and the number of 696 working days in this period, we have an average of 12 issues per day. Thus, approximately 13 minutes (780 seconds) are spent per issue in the manual assignment, while the proposed assignment system takes only 161.55 milliseconds per issue, including HTTP requests' processing time. This average time was collected executing the automated assignment of 100 issues. However, we need to consider that only 81.63\% of these responses will be correct on average (see Equation \ref{eq:finalAcc}). 

In that way, considering a day with 12 new issues, nine would be classified correctly, taking 1.45 seconds $(9\:issues \times 161.55\:milliseconds/issue) / 1000$ and three would be classified incorrectly taking 2,340 seconds to correct this misassignment ($3\:issues \times 780\:seconds/issue$), resulting in  2,341.45 seconds or approximately 39 minutes. \myhl{Here, it is worth mentioning two points: first, the average time spent to do the manual assignment (13 minutes) is high due to problems in the usability of the issue tracking system. This system does not have notifications for new issues, and it requires users to update their sessions many times a day. These problems improve the benefits of our proposal. Second, according to the domain experts, wrong assignments do not increase the assignment cost and complexity. So, this point was not included in our saving analysis.}

We can then answer the RQ5 stating that there is a reduction of 75.62\% in the time spent on this activity when using the automated strategy, even considering the errors. The time savings are approximately 40 hours per month, which represents a significant achievement for the teams.

\section{Lessons Learned}
\label{sec:lessons}
This study was not conducted to propose another algorithm to deal with the issue assignment. Instead, our goal is to present a report about how existent algorithms and processes can be applied in a real project and what we could learn from this experience. Thus, four lessons learned are discussed.

\textbf{A well-defined project, an iterative process, and an open communication channel are essential}: these three items are crucial for software projects, especially in an industry-academia partnership, to enable a productive experience exchange. In our case, a research project that is focused on improving the company's processes, the CRISP-DM was decisive for achieving the project's goals because it has well-defined steps and enables quick iterations focused on the practical problem. Also, the periodic meetings contributed to understanding the company's particularities and needs.

\textbf{Certain flexibility to add features to facilitate the use of new proposed solutions is a good practice}: the relationship between industry and academia faces several challenges. Among them, we can highlight the difficulty of incorporating the knowledge and integrating the new solutions to the already existing context \cite{garousi2016challenges}. In our case, it was only requested a Web service accessible through an API for later integration with the company's issue tracking system. However, due to the aforementioned challenges, it was decided to develop a series of additional functionalities (with graphical user interfaces) to support the automated assignment service's maintenance and increase the team's confidence in the transition period (from manual assignment to auto-assignment). Among these features, it is worth mentioning: i) batch classification in which it is possible to export the results to CSV files, allowing team leaders to test the tool before its integration; and ii) training and report center that automate the models' creation from data that can be inserted by the leaders or obtained through the issue tracking API. These features increased managers' confidence as they understood that there is a continuity plan without additional effort for maintaining the new tool.

\textbf{The deterioration of the models should be investigated}: the models' deterioration is natural, especially when significant changes can occur in data over time. Thus, it is relevant to investigate this problem to create strategies to overcome it. In the literature, there are works proposing online methods to detect the accuracy decrease, such as monitoring changes in the assignee after the automated assignment \cite{aktas2020automated}. In our case, this strategy was not adequate because it is common to change the issues' assignee after a careful bug analysis. Thus, we opted for an investigation with offline experiments using sliding windows for training and testing. This was essential for choosing the retraining frequency and historical data window, facilitating the service maintenance.

\textbf{Automated assignment can deliver value to the client even when the accuracy is not so high}: when faced with a low accuracy such as those obtained by the Random Forest classifier for sub-teams $ST_ {4}, ST_ {5}, ST_ {6}$, the first idea that comes up is to conduct more in-depth investigations to find new techniques that can improve this result. However, this decision may delay the delivery of preliminary results that already bring gains for the industry. In our case, even with a success probability of 81.63\%, which can be seen as a low probability for a problem that has only six classes, the effort saved (79.98\%) with these current models is still significant. Understanding this aspect allows the research team to be agile in delivering results throughout the process execution and with improvements made incrementally. This attitude avoids long waits for practical results and a lack of motivation.

\myhl{We summarize our main insights taken from this industry-academia collaboration as follows.
It is important to design the research project to adapt scientific rigor to the company's spirit, prioritizing processes that deliver value frequently.
Establish a trust relationship to avoid a sense of threat in employees.
The research team's flexibility to develop additional features with good usability is an appealing way to engage managers and domain experts.
It is also essential to investigate and document strategies for maintaining all artifacts after the project's end.
Finally, even below optimal results can generate value for the company depending on the context and the research scope.
Thus, we do not recommend delaying deliveries to try out too complex methods as the results can be just slightly better.}

\section{Conclusions}
\label{sec:conclusions}
This paper presented an industrial report about using machine learning to optimize the issue assignment in a large electronic company. In our case, the problem was to build intelligent models that could correctly distribute new issues according to the teams' responsibility, given historical data.

As a result of this experience, we highlight the importance of a well-defined project, an iterative process, and an open communication channel between researchers and practitioners. It is also crucial to develop additional features to support service maintenance and investigate the models' deterioration to define the suitable retraining frequency. Finally, \myhl{depending on the context}, even when accuracy is not so high, the automated assignment can represent a significant achievement for the teams due to the time/effort saved.

For future work, we intend to investigate the aspects used by the leaders to perform the manual assignment. With this information, we can create an expert system and integrate it with our learning models. Other interesting points are balancing for textual data and trying deep learning models.

\section*{Data Availability}
All data and scripts used in this work are protected by a non-disclosure agreement.

\section*{Acknowledgments}

We would like to thank CNPq for the Productivity Scholarship of Rossana M. C. Andrade DT-2 ($N^{o}$ 315543 / 2018-3) and LGE Brazil for supporting this research under the Brazilian Informatics Law ($N^{o}$ 10.176 of 1/11/2001) incentives.

\end{document}